\documentclass[12pt]{iopart}
\usepackage{graphicx}

\begin{document}

\def\Journal#1#2#3#4{{#1} {\bf #2}, #3 (#4)}

\def\NCA{Nuovo Cimento}
\def\NIM{Nucl. Instr. Meth.}
\def\NIMA{{Nucl. Instr. Meth.} A}
\def\NPB{{Nucl. Phys.} B}
\def\NPA{{Nucl. Phys.} A}
\def\PLB{{Phys. Lett.}  B}
\def\PRL{Phys. Rev. Lett.}
\def\PRC{{Phys. Rev.} C}
\def\PRD{{Phys. Rev.} D}
\def\ZPC{{Z. Phys.} C}
\def\JPG{{J. Phys.} G}
\def\CPC{Comput. Phys. Commun.}
\def\EPJ{{Eur. Phys. J.} C}

\title[$\pi, K, p$ and $\bar{p}$ production from
Au+Au collisions at $\sqrt{s_{_{NN}}} = 62.4$ GeV]{$\pi, K, p$ and
$\bar{p}$ production from Au+Au collisions at $\sqrt{s_{_{NN}}} =
62.4$ GeV}

\author{Lijuan Ruan\footnote[1]{This work is supported in part by NSFC (10155002).}\footnote[2]{rlj@mail.ustc.edu.cn}
(for the STAR Collaboration\footnote[3]{For the full author list
and acknowledgements see Appendix ``Collaborations'' in this
volume.})}
\address{Dept. of Modern Physics, University of Science and Technology of China, Hefei, Anhui, China, 230026}
\begin{abstract}
The preliminary results of $\pi^{\pm}, K^{\pm}, p$ and $\bar{p}$
spectra are reported from Au+Au collisions at $\sqrt{s_{_{NN}}} =
62.4$ GeV. Particle identification is from the Time Projection
Chamber and Time-of-Flight system at STAR.  The nuclear
modification factor $R_{CP}$ for mesons ($\pi^{\pm}, K^{\pm}$) and
baryons ($p, \bar{p}$) will also be discussed.
\end{abstract}
\vspace{-0.35cm}
In central 200 GeV Au+Au collisions at RHIC, hadron production at
high $p_T$ ($p_T>$5 GeV/c) has been observed to be suppressed
relative to that from p+p collisions scaled by the number of
binary collisions and the measured nuclear modification factor is
significantly below unity~\cite{starhighpt,phenixhighpt}. This
suppression has been interpreted as energy loss of the energetic
partons traversing the produced hot and dense
medium~\cite{jetquench}. At intermediate $p_{T}$ ($2<p_{T}<5$
GeV/c), the degree of suppression depends on particle species. The
nuclear modification factors of baryons (protons and lambdas) are
significantly higher than those of mesons (pions, kaons)
~\cite{starv2raa,phenixpid}, which can be explained by coalescence
of constituent quarks at hadronization from a collective partonic
system~\cite{1}. In Au+Au collisions at 62 GeV, the bulk
properties created was expected to be different from those at 200
GeV. The smaller gluon density was expected~\cite{vitevwang62}. In
such an environment, we'd like to study the energy loss behavior
and the particle composition at intermediate and high $p_{T}$. In
this paper, the particle spectra and their nuclear modification
factors of $\pi,K,p$ and $\bar{p}$ will be reported. Also
presented are the ratios of protons over pions. \vspace{-0.5cm}
\section{Analysis methods and results}
\vspace{-0.35cm}
\begin{figure}[h] \centerline{
\includegraphics[width=0.4\textwidth]
{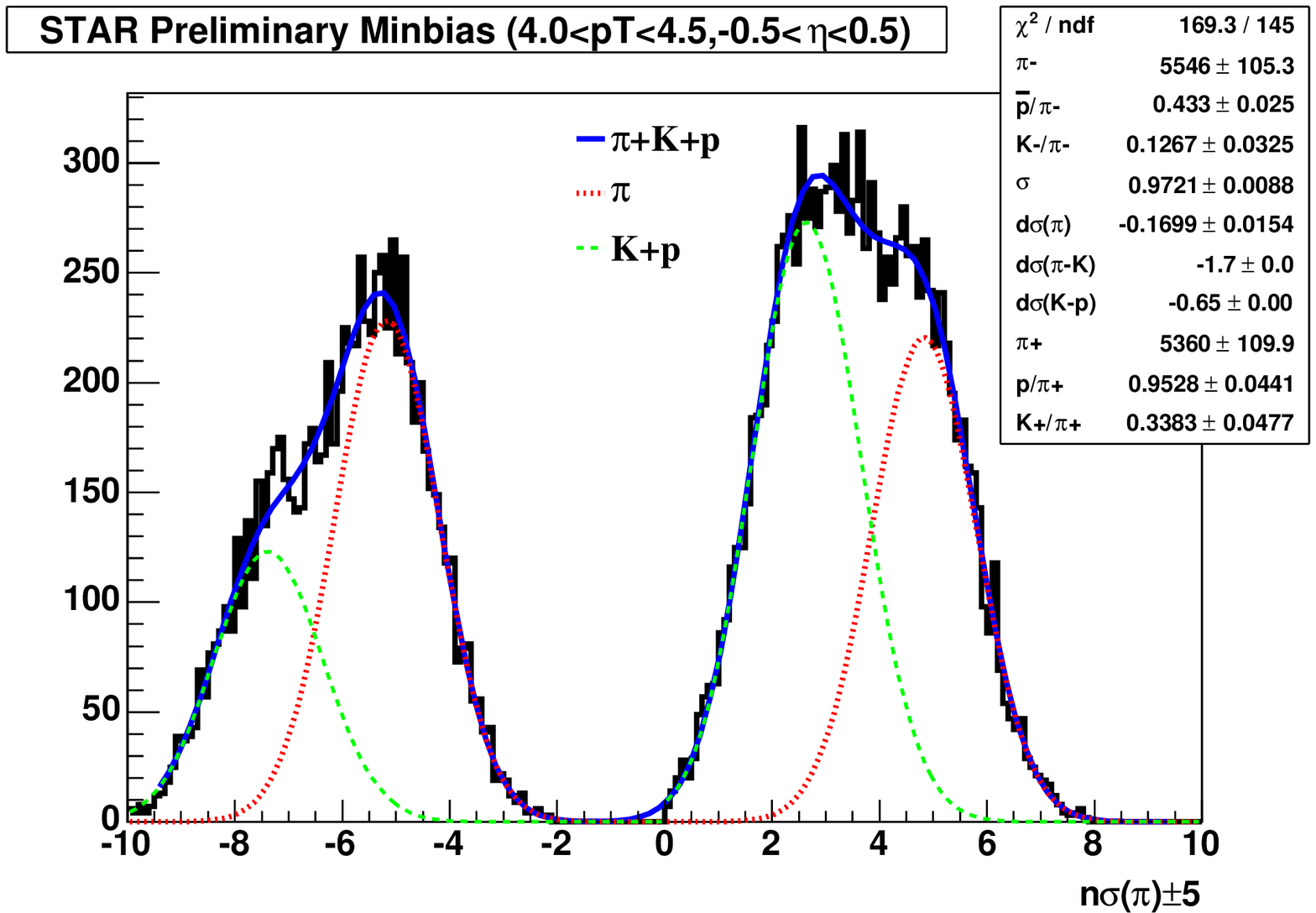}\includegraphics
[width=0.4\textwidth]{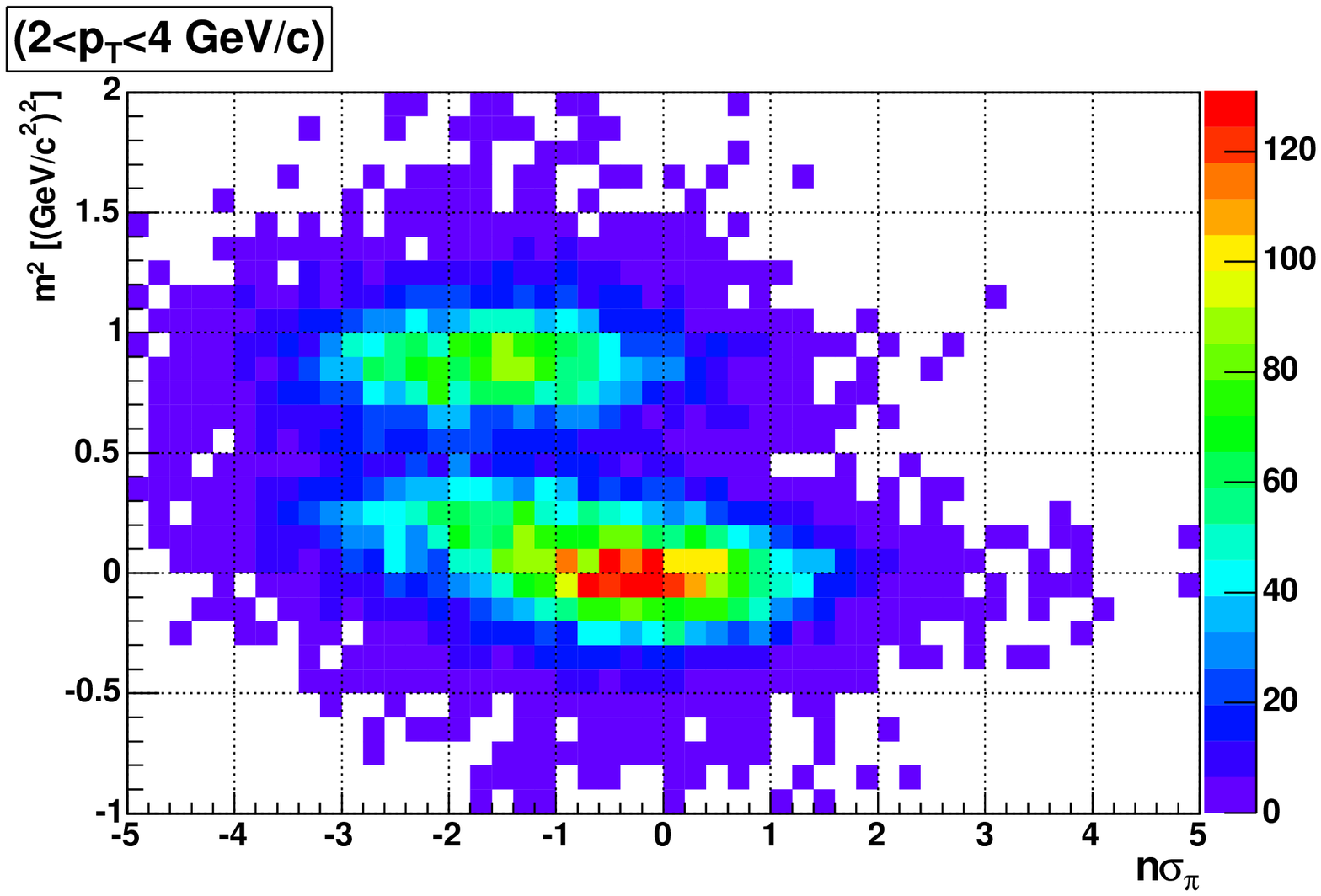}} \vspace{-0.35cm}

\caption[]{(left) dE/dx distribution normalized by pion dE/dx and
offset by $\pm5$ for positive and negative charged particles at
$4<p_T<4.5$ GeV/c, respectively. The distribution is from
minimum-bias Au+Au collisions at 62.4 GeV~\cite{zhangbu62}.
(right) $m^2$ versus $n\sigma_{\pi}$ at $2<p_T<4$ GeV/c in
minimum-bias Au+Au collisions at 62.4 GeV.}
\vspace{-0.6cm}\label{fig1}
\end{figure}

At STAR, charged particles can be identified up to $p_{T}\sim1.1$
GeV/c by measuring their ionization energy loss ({\it dE/dx}) in
the Time Projection Chamber (TPC)~\cite{startpctof}. The {\it
dE/dx} resolution was calibrated to be better than 8\% in 62.4 GeV
Au+Au collisions for tracks with 70 cm in length inside the TPC.
At momentum $p>$ 3 GeV/c, the {\it dE/dx} of $\pi^{\pm}$ has a
$\sim 2\sigma$ separation from those of $K^{\pm}$ and $p(\bar{p})$
due to the relativistic rise of pion {\it dE/dx}. Thus charged
pions can be identified at $3<p_{T}<10$ GeV/c~\cite{zhangbu62}.
The {\it dE/dx} measurement in Fig.~\ref{fig1} (left) uses a
normalized {\it dE/dx}:
$n\sigma_{X}^{Y}=ln((dE/dx)^{Y}/I_{0X})/\sigma_{X}$, where $X,Y$
can be $e^{\pm},\pi^{\pm},K^{\pm}$ or $p(\bar{p})$,$(dE/dx)^{Y}$
is the measured {\it dE/dx} of particle $Y$, $I_{0X}$ is the
expected $dE/dx$ of particle $X$, and $\sigma_{X}$ is the $dE/dx$
resolution of particle $X$. With a perfect calibration, the
$n\sigma_{\pi}^{\pi}$ distribution should be a normal Gaussian
while $n\sigma_{\pi}^{p},n\sigma_{\pi}^{K}$ should be a Gaussian
peaked at negative value. The method was introduced
in~\cite{zhangbu62}. A prototype multi-gap resistive plate chamber
time-of-flight system (TOFr)~\cite{startpctof} has been installed
since 2003 with the coverage $-1\!<\!\eta\!<\!0$ in
pseudorapidity. With intrinsic timing resolution 85
ps~\cite{starcronin}, it extends particle identification up to
$p_{T}\sim3$ GeV/c for $p$ and $\bar{p}$ and 1.6 GeV/c for
$\pi^{\pm}$ and $K^{\pm}$. By the combination of
$m^{2}=p^{2}(1/\beta^{2}-1)$ from TOFr and
$n\sigma_{\pi},n\sigma_{p}$ information from TPC, $\pi^{\pm}$, $p$
and $\bar{p}$ can be identified up to $p_{T}\sim5$ GeV/c and
$K^{\pm}$ can be identified to at least 3 GeV/c, where $\beta$ is
the velocity. This method has been introduced in~\cite{ming62}.
Fig.~\ref{fig1} (right) shows $m^{2}$ versus $n\sigma_{\pi}$.

A total of 6.8 million events after vertex cut $|Vz|<$30 cm were
used for the analysis, where $Vz$ is the z position of the vertex.
Centrality tagging of Au+Au collisions was based on the charged
particle multiplicity in $-0.5<\eta<0.5$, measured by the TPC. The
minimum-bias (0-80\%) events were divided into four centralities:
most central $10\%$, $10-20\%$, $20-40\%$ and $40-80\%$ of the
hadronic cross section.

From TOFr, the raw yields of $\pi^{\pm}$, $K^{\pm}$, $p$ and
$\bar{p}$ are obtained from Gaussian fits to the distributions in
$m^{2}=p^{2}(1/\beta^{2}-1)$ in each $p_{T}$ bin w/o $n\sigma$
cut. Acceptance and efficiency were studied by Monte Carlo
simulations and by matching TPC track and TOFr hits in real data.
From TPC, the raw yields of $\pi^{\pm}$ were extracted from
$dE/dx$ distribution. The efficiency due to the additional
$n\sigma$ cut was also taken into account. Weak-decay feeddown
(e.g. $K_{s}^{0}\rightarrow\pi^{+}\pi^{-}$) to pions was not
corrected for, which was estimated to contribute $\sim$ 12\% at
$p_{T}<$1 GeV/c and $\sim$ 5\% at higher $p_{T}$ to pion
yields~\cite{starcronin}. Inclusive $p$ and $\bar{p}$ production
is presented without hyperon feeddown correction either. $p$ and
$\bar{p}$ from hyperon decays have the same detection efficiency
as primary $p$ and $\bar{p}$~\cite{olgaprl,antiproton} and the
contributions to the inclusive $p$ and $\bar{p}$ yield range from
$\sim$20\% to $\sim$40\% from p+p, d+Au to Au+Au
collisions~\cite{starcronin,olgaprl,antiproton}.

\begin{figure}[h] \vspace{-0.3cm} \centerline{
\includegraphics[width=0.3\textwidth]
{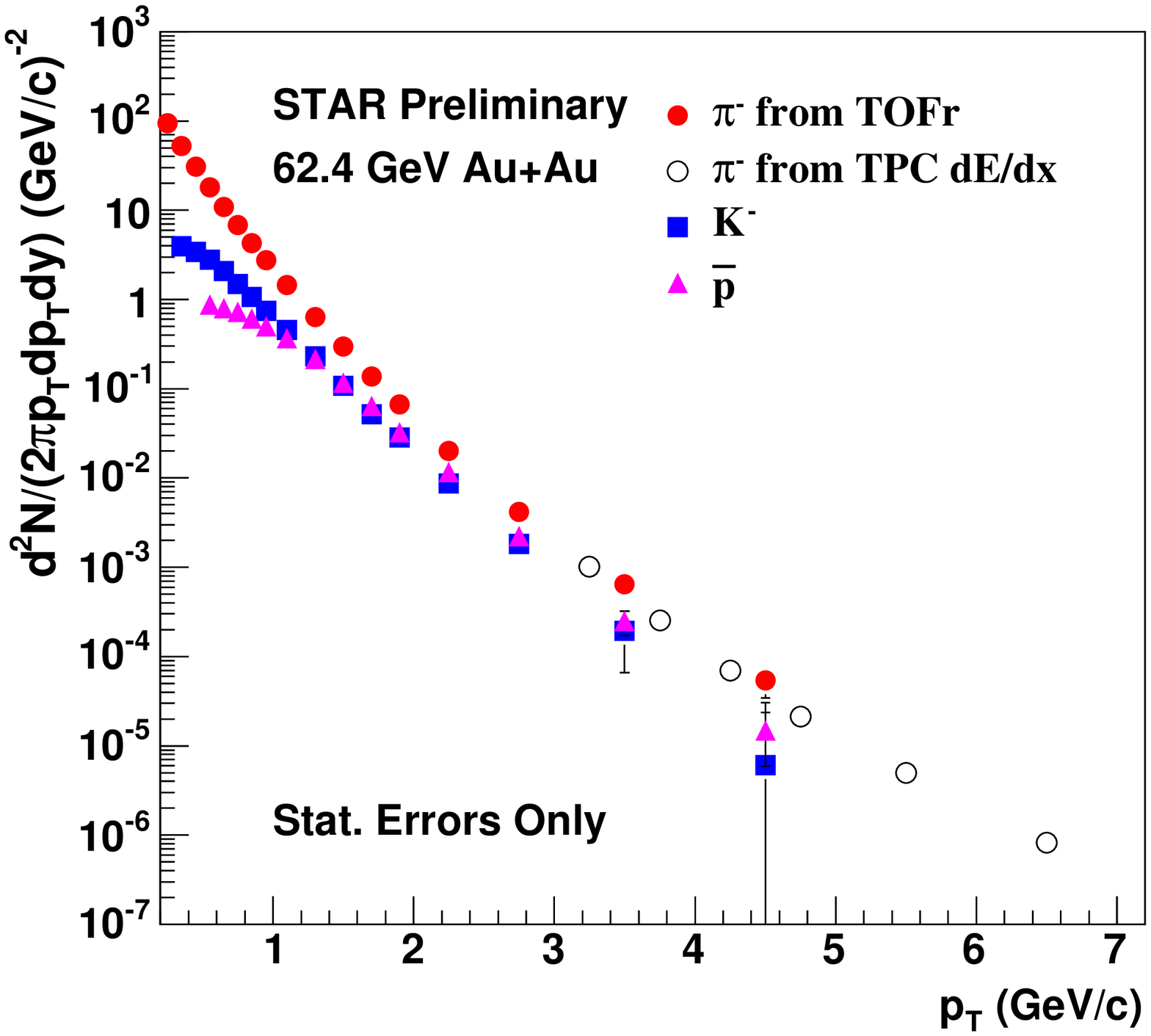}\includegraphics
[width=0.3\textwidth]{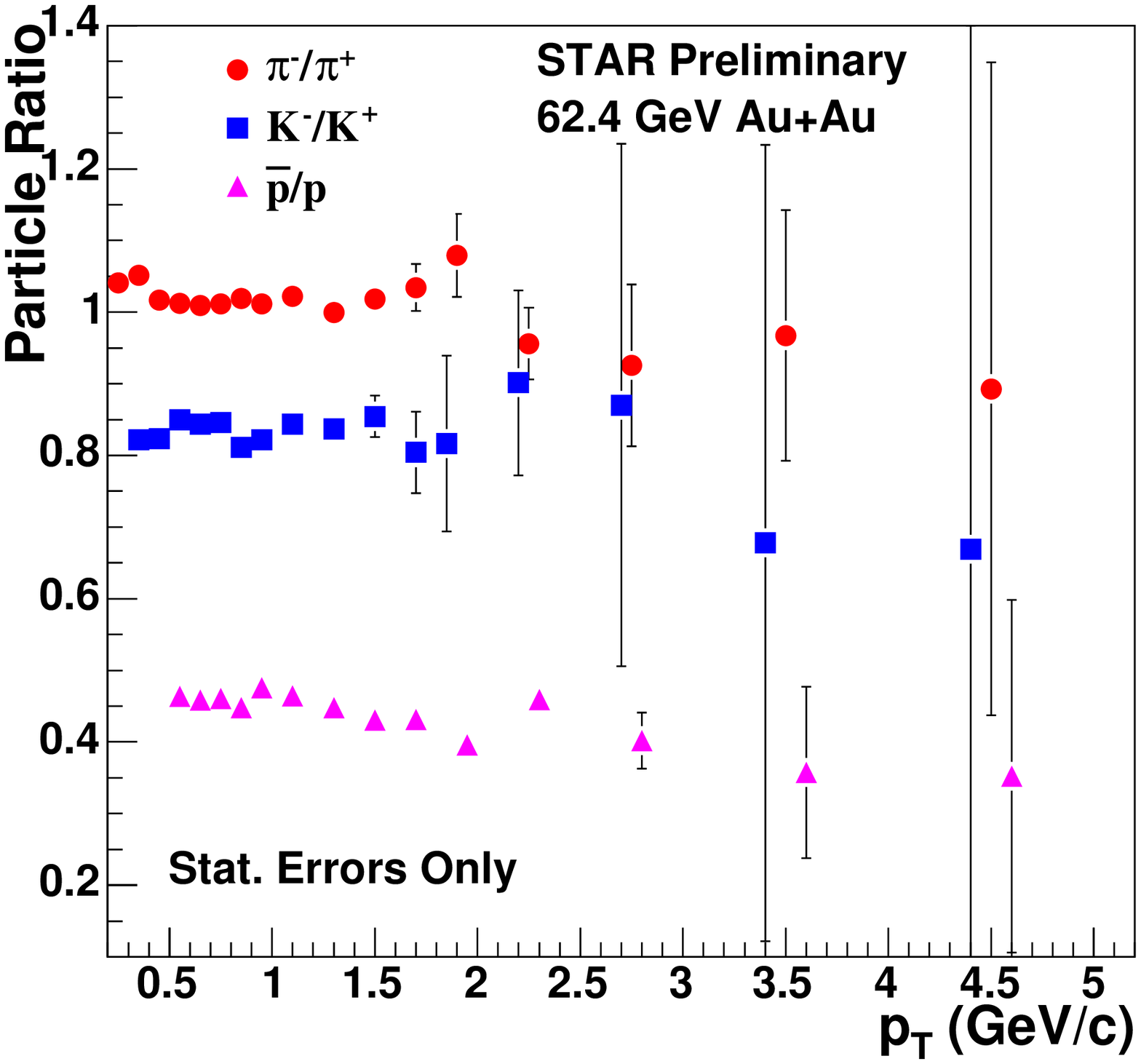}
\includegraphics[width=0.3\textwidth]
{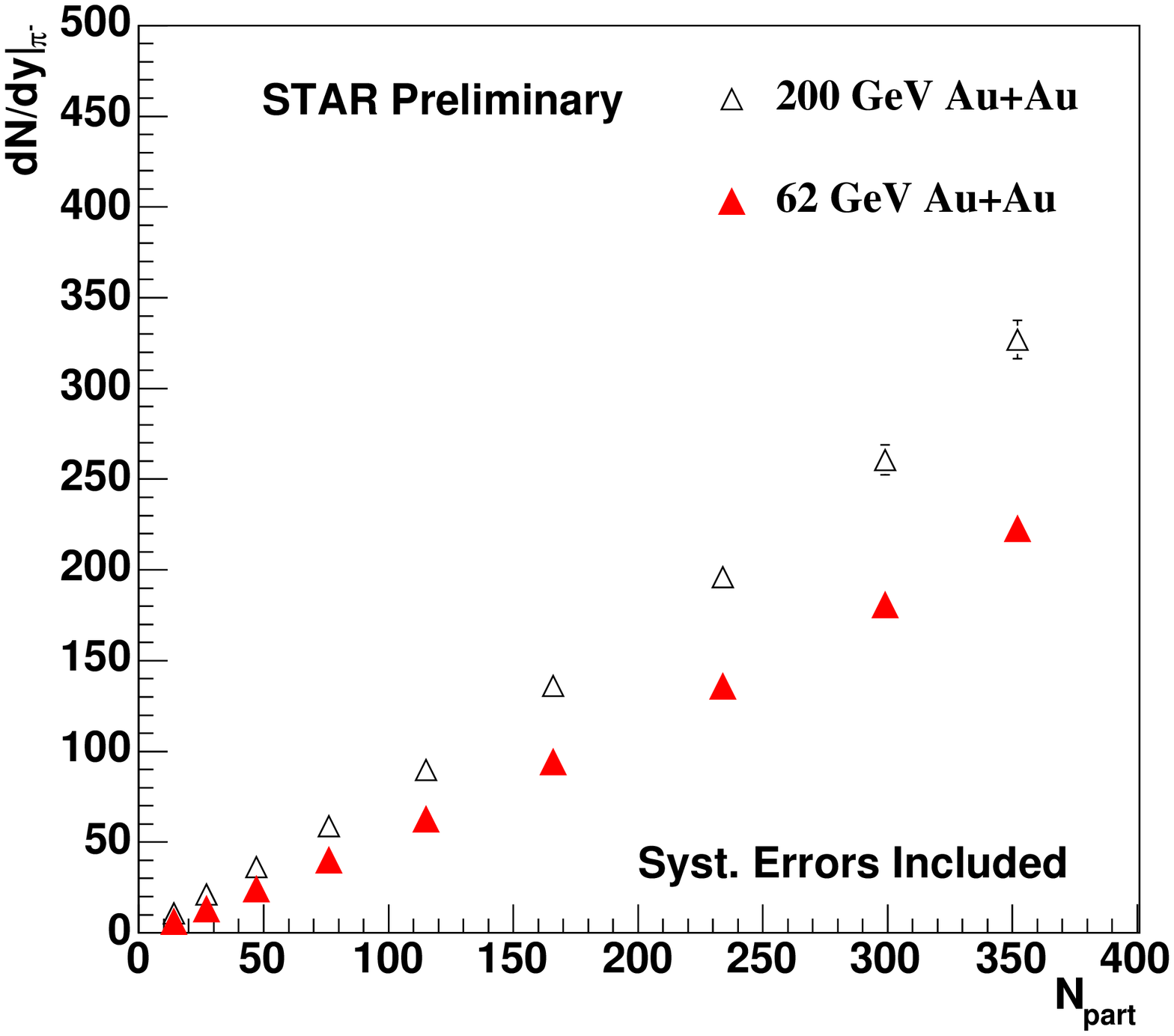} } \vspace{-0.35cm}
\caption[]{(left) The spectra of $\pi^{-},K^{-}$ and $\bar{p}$
versus $p_T$. (middle) $\pi^{-}/\pi^{+}, K^{-}/K^{+}$ and
$\bar{p}/p$ ratios versus $p_T$. (right) The $\pi^{-}$ rapidity
density versus $N_{part}$.} \vspace{-0.60cm}\label{fig2}
\end{figure}
The invariant yields $d^2N/2\pi p_Tdp_Tdy$ of $\pi^{-},K^{-}$ and
$\bar{p}$ at mid-rapidity, after the efficiency correction, are
shown as symbols in Fig. 2 (left) for 62.4 GeV minimum bias Au+Au
collisions. In the overlapping $p_{T}$ region, the results from
TOFr and from TPC are consistent. Fig. 2 (middle) shows the
anti-particle to particle ratios versus $p_{T}$. Within the
errors, $\pi^{-}/\pi^{+}, K^{-}/K^{+}$ and $\bar{p}/p$ are flat
with $p_T$. The average anti-particle to particle ratios were
obtained with a fit of constant value:
$\pi^{-}/\pi^{+}=1.02\pm0.01$,$K^{-}/K^{+}=0.84\pm0.01$ and
$\bar{p}/p=0.46\pm0.01$. The errors are statistical. The
systematical errors are similar to those presented
at~\cite{starcronin,olgaprl}. At 200 GeV, $\bar{p}/p=0.77\pm0.05$.
The decrease of $\bar{p}/p$ ratio at 62 GeV indicates the increase
of the baryon chemical potential. The mid-rapidity yield $dN/dy$
of $\pi^{-}$ was extracted with a Bose-Einstein
fit~\cite{olgaprl}. Fig. 2 (right) shows the $dN/dy$ of $\pi^{-}$
versus $N_{part}$ in Au+Au collisions at 62 and 200
GeV~\cite{olgaprl}. The $dN/dy$ at 62 GeV is a factor of $\sim$1.5
smaller than that at 200 GeV~\cite{olgaprl}.

\begin{figure}[h] \vspace{-0.5cm}\centerline{
\includegraphics[width=0.3\textwidth]
{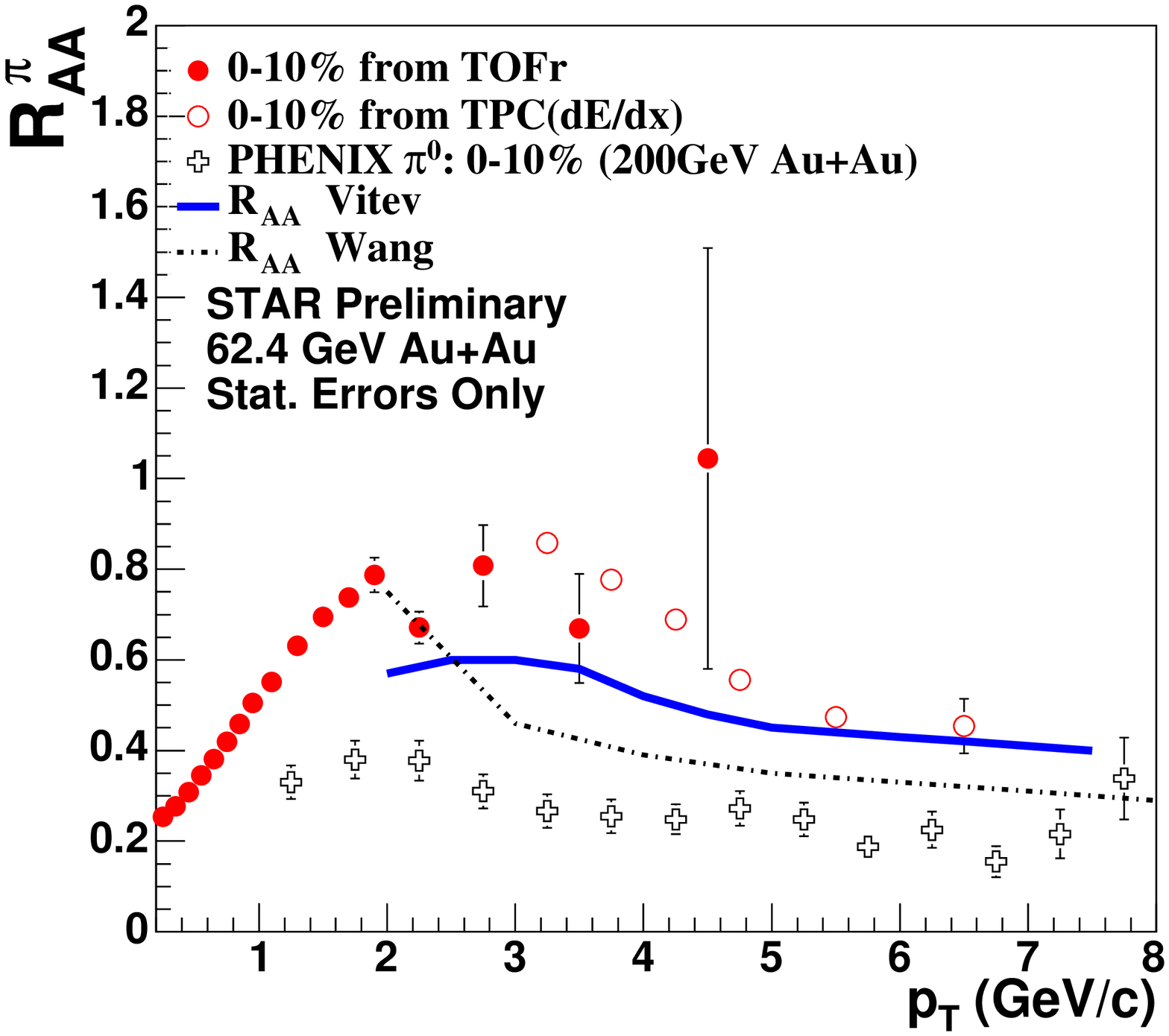}\includegraphics
[width=0.3\textwidth]{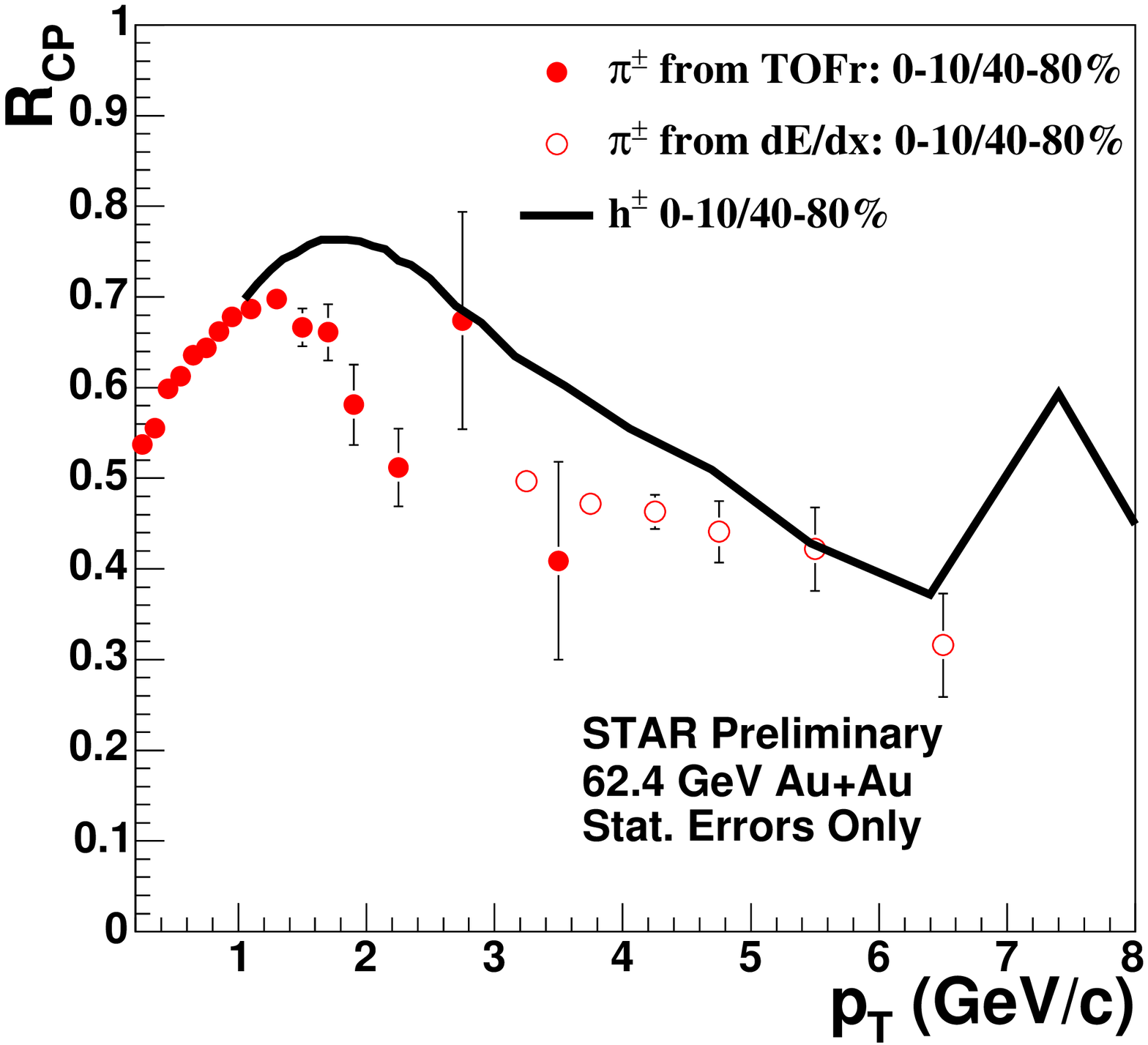}
\includegraphics[width=0.3\textwidth]
{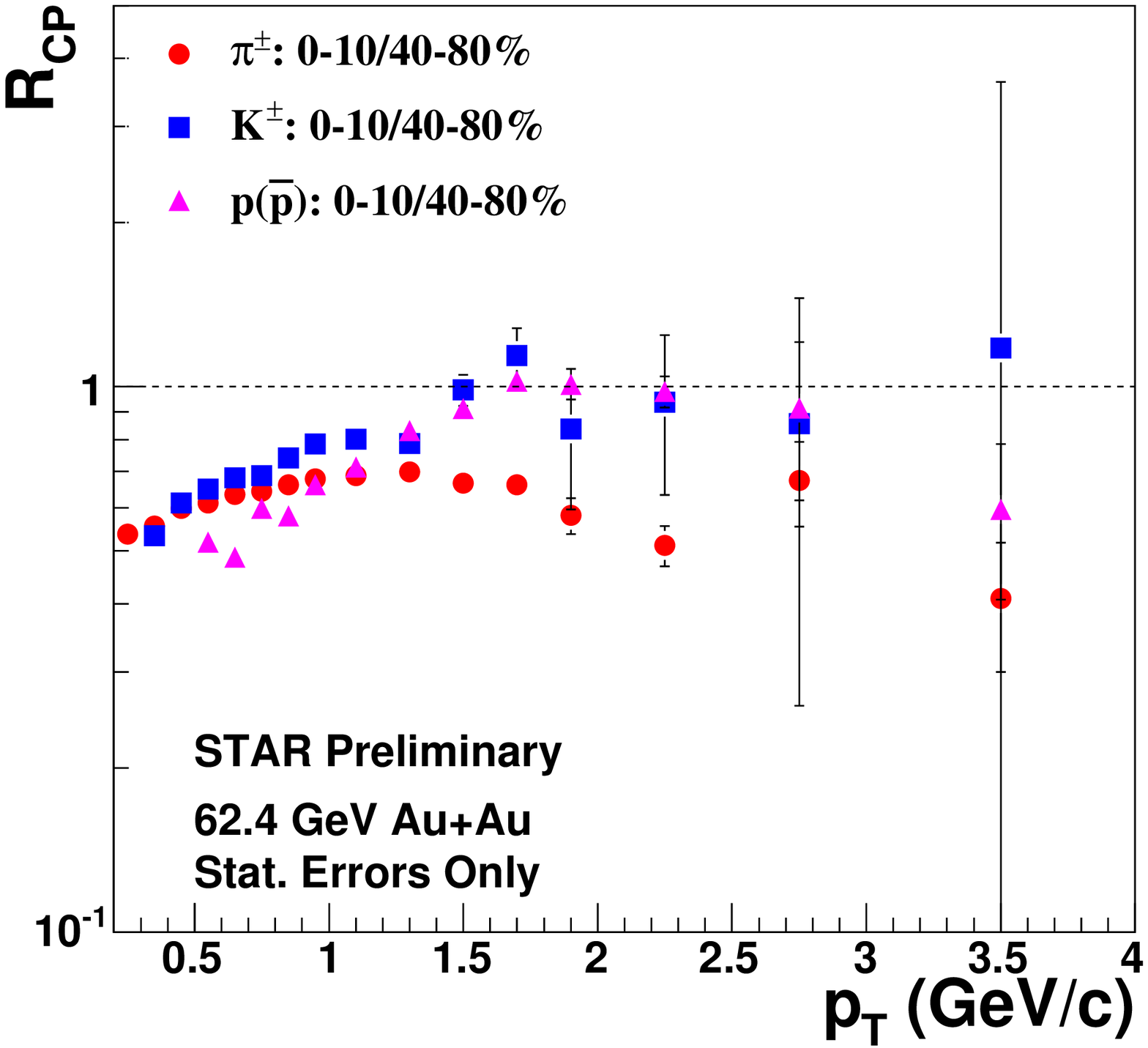} } \vspace{-0.35cm} \caption[]{(left)
$R_{AA}$ of $\pi^{\pm}$ versus $p_T$ in Au+Au collisions at 62.4
GeV. Also shown are the two curves from theoretical predictions
for 62.4 GeV~\cite{vitevwang62}. Reference spectrum of p+p is from
ISR~\cite{piISRphenixPara}. (middle) $R_{CP}$ of $\pi^{\pm}$ and
$h$. (right) $R_{CP}$ of $\pi^{\pm}, K^{\pm}$ and $p+\bar{p}$. }
\label{fig3}
\end{figure}

\vspace{-0.5cm} Nuclear effects on hadron production in Au+Au
collisions were measured through comparison to the p+p spectrum
scaled by the number of underlying nucleon-nucleon inelastic
collisions ($N_{bin}$). Fig. 3 (left) shows the $R_{AA}$ versus
$p_{T}$ from our measurement in most central 0-10\% Au+Au
collisions at 62.4 GeV. Also shown in this plot is the $R_{AA}$ in
most central 0-10\% Au+Au collisions at 200 GeV from PHENIX
measurement~\cite{phenixhighpt}.  It's evident that the
suppression exists at 62.4 GeV and that the magnitude of
suppression is smaller than that at 200 GeV. Fig. 3 (middle) shows
$R_{CP}$ of charged hadron ($h$) and $\pi^{\pm}$ from 62.4 GeV
Au+Au collisions. $R_{CP}$ of $h$~\cite{starhadron62} is larger
than that of $\pi^{\pm}$ at $2<p_T<5$ GeV/c and approaches that of
$\pi^{\pm}$ at $5<p_T<7$ GeV/c. This may indicate the
disappearance of particle-species dependence of nuclear
modification factor at high $p_T$. Fig. 3 (right) shows $R_{CP}$
of $\pi^{\pm}, K^{\pm}$ and $p+\bar{p}$ from 62.4 GeV Au+Au
collisions. $R_{CP}$ of protons seems to follow $N_{bin}$ scaling
at intermediate $p_T$ and be larger than those of pions. The
statistic for kaons is too poor to address physics issues.

Fig. 4 shows the ratios of $p/\pi^{+}$ and $\bar{p}/\pi^{-}$ as a
function of $p_T$ in Au+Au collision at 62.4 GeV. The $p/\pi^{+}$
and $\bar{p}/\pi^{-}$ ratios are observed to be a factor of
$2\sim3$ higher than those in p+p collisions at similar energy at
$2<p_{T}<4$ GeV/c. $p/\pi^{+}$ and $\bar{p}/\pi^{-}$ ratios reach
maximum at around 2 GeV/c and seem to decrease with increasing
$p_T$ at higher $p_T$. This is consistent with the trends of
$R_{CP}$ of $h$ and $\pi$. At $p_{T}\sim5$ GeV/c, $p/\pi^{+}$ and
$\bar{p}/\pi^{-}$ ratios in 62 GeV Au+Au collisions are close to
those in p+p collisions~\cite{ISRpp1975}. This may indicate that
fragmentation mechanism dominates above this $p_{T}$ region.

\begin{figure}[h] \vspace{-0.5cm} \centerline{
\includegraphics[width=0.3\textwidth]
{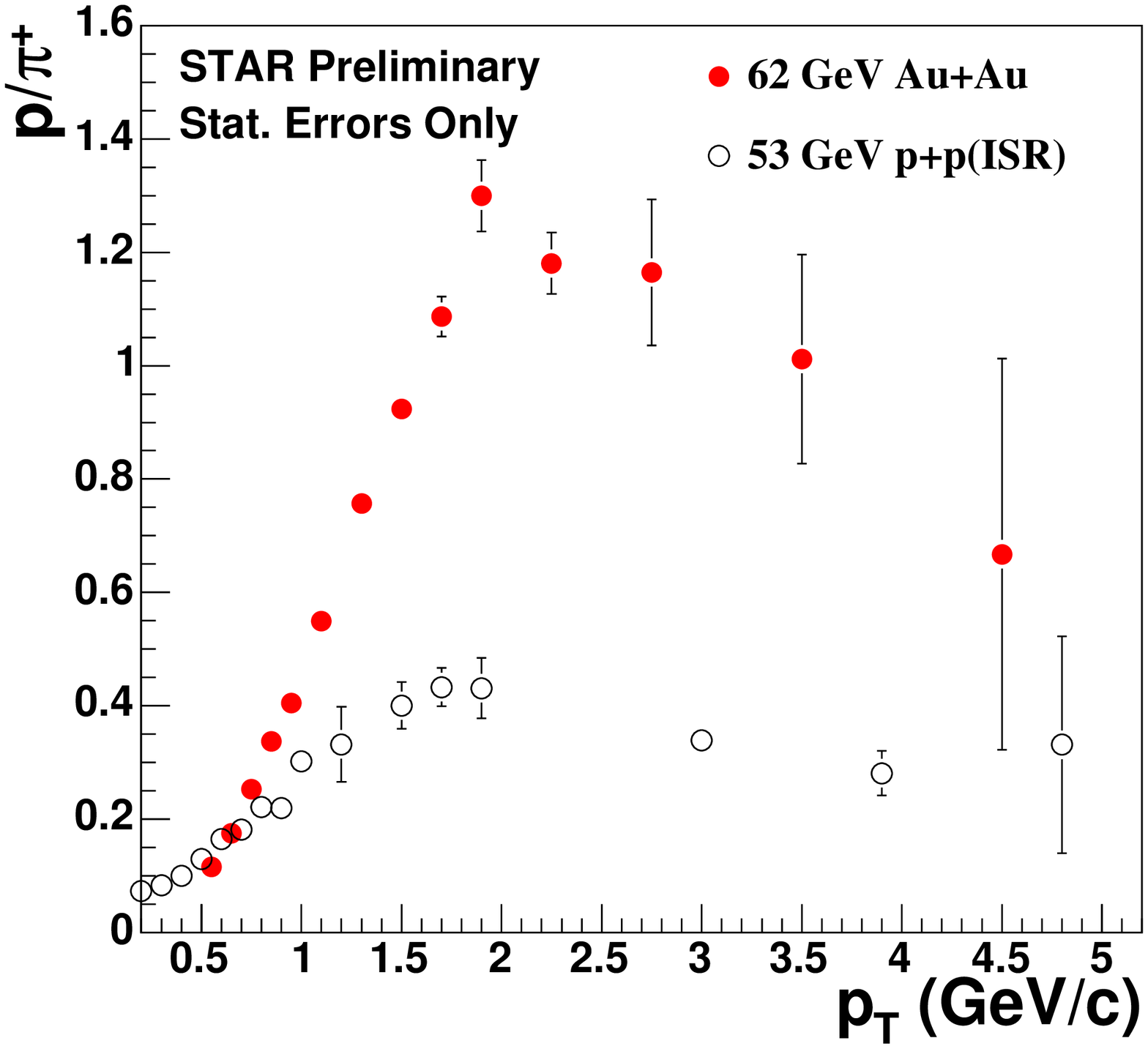}\includegraphics
[width=0.3\textwidth]{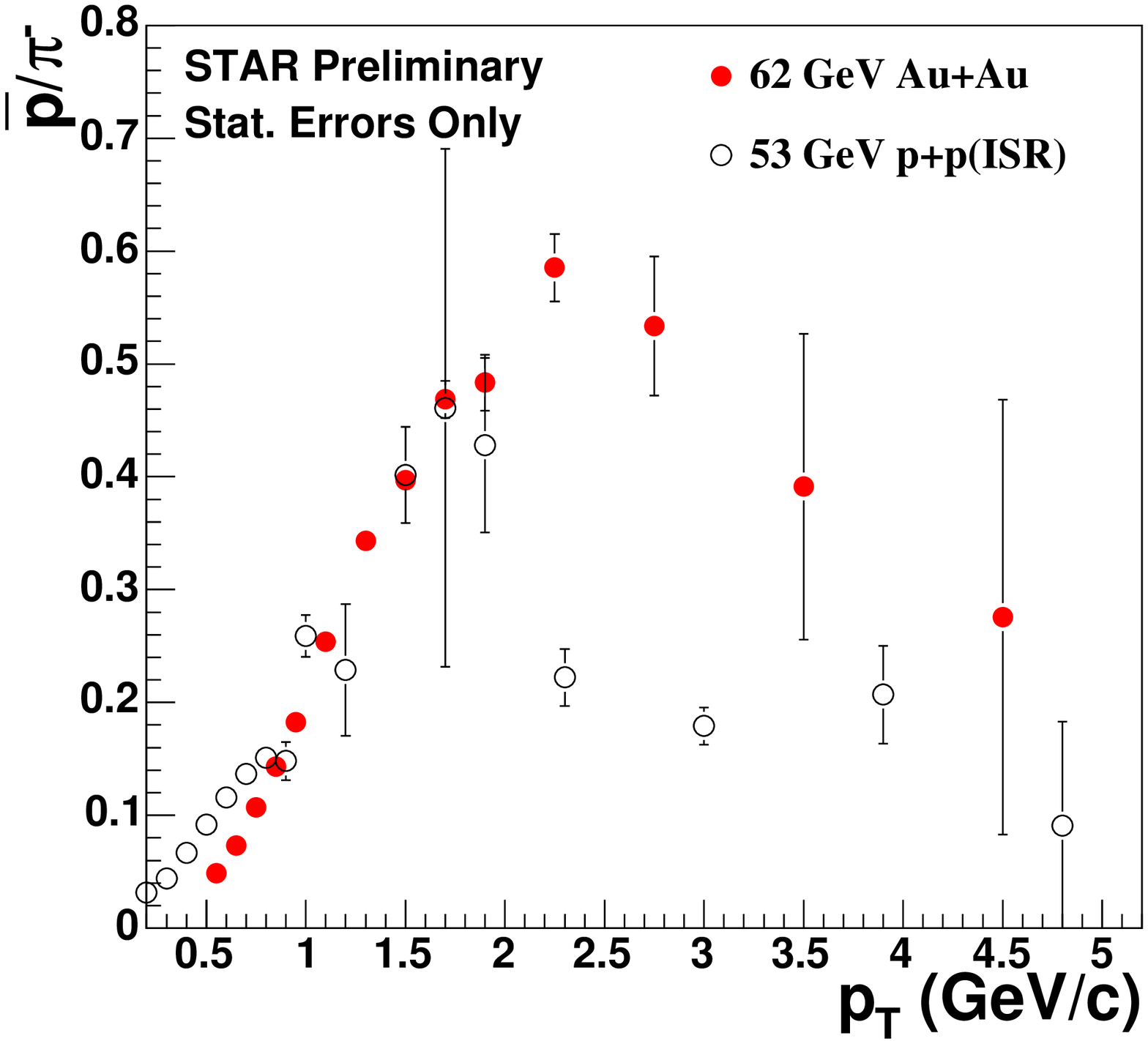}} \vspace{-0.35cm}
\caption[]{The ratios of $p/\pi^{+}$ (left) and $\bar{p}/\pi^{-}$
(right) as a function of $p_T$ in Au+Au collision at 62.4 GeV.
Also shown are the ratios in p+p collisions at 53
GeV~\cite{ISRpp1975}.} \label{fig4}
\end{figure}

\vspace{-1.2cm}
\section{Summary}\vspace{-0.35cm}
We report the STAR preliminary results of $\pi^{\pm}, K^{\pm}, p$
and $\bar{p}$ spectra from 62.4 GeV Au+Au collisions. $\pi^{\pm}$
and $p(\bar{p})$ were identified up to $p_T$ $\sim$7 GeV/c and 5
GeV/c, respectively. At this beam energy, $\bar{p}/p=0.46\pm0.01$.
A significant suppression for pions is observed for the most
central collisions, but the effect is weaker than that observed in
200 GeV central Au+Au collisions~\cite{phenixhighpt} at $p_T<$7
GeV/c. At intermediate $p_T$, the nuclear modification factor
$R_{CP}$ of $h$ is 20\% higher than that of $\pi$, $R_{CP}$ of
$p+\bar{p}$ is larger than that of $\pi$, and the ratios of
$p/\pi^{+}$ and $\bar{p}/\pi^{-}$ are a factor of 2$\sim$3 higher
than those in p+p collisions at similar energies~\cite{ISRpp1975}.


\vspace{-0.6cm}

\section*{References}
\vspace{-0.35cm}
\begin{tiny}
 \end{tiny}
\end{document}